\documentclass{article}

\usepackage{arxiv}

\usepackage[utf8]{inputenc}
\usepackage[T1]{fontenc}
\usepackage{lmodern}

\usepackage{hyperref}
\hypersetup{hypertexnames=false}
\usepackage{xurl}
\usepackage{booktabs}
\usepackage{amsfonts}
\usepackage{amsmath}
\usepackage{amssymb}
\usepackage{mathtools}

\usepackage{graphicx}
\usepackage{tikz}
\usetikzlibrary{arrows.meta,positioning,fit,backgrounds,calc}
\usepackage{microtype}

\usepackage{algorithm}
\usepackage{algpseudocode}

\usepackage{enumitem}
\usepackage{listings}
\usepackage{multirow}
\usepackage{tabularx}

\usepackage[numbers,compress]{natbib}
\usepackage{doi}
\usepackage{placeins}

\newcolumntype{Y}{>{\raggedright\arraybackslash}X}
\definecolor{layerGoal}{HTML}{DCEBFF}
\definecolor{layerDialogue}{HTML}{E8F5E9}
\definecolor{layerExec}{HTML}{FFF3D6}
\definecolor{layerPolicy}{HTML}{F3E5F5}
\definecolor{lineDark}{HTML}{34495E}
\definecolor{textDark}{HTML}{1F2933}
\definecolor{conversationBg}{HTML}{F8FAFC}
\definecolor{conversationUser}{HTML}{1D4ED8}
\definecolor{conversationSystem}{HTML}{047857}
\definecolor{conversationNote}{HTML}{6B7280}
\definecolor{codeBg}{HTML}{F8FAFC}
\definecolor{codeRule}{HTML}{CBD5E1}
\definecolor{codeKeyword}{HTML}{1D4ED8}
\definecolor{codeComment}{HTML}{64748B}

\lstdefinestyle{runtime}{
	basicstyle=\ttfamily\footnotesize,
	breaklines=true,
	columns=fullflexible,
	keepspaces=true,
	showstringspaces=false,
	frame=single,
	backgroundcolor=\color{codeBg},
	rulecolor=\color{codeRule},
	framerule=0.45pt,
	framesep=6pt,
	xleftmargin=0.75em,
	xrightmargin=0.75em,
	aboveskip=0.8em,
	belowskip=0.8em
}

\lstdefinestyle{python}{
	style=runtime,
	language=Python,
	keywordstyle=\color{codeKeyword}\bfseries,
	commentstyle=\color{codeComment}\itshape,
	stringstyle=\color{conversationSystem}
}

\lstdefinestyle{conversation}{
	basicstyle=\ttfamily\small,
	breaklines=true,
	columns=fullflexible,
	keepspaces=true,
	frame=single,
	backgroundcolor=\color{conversationBg},
	rulecolor=\color{lineDark!45},
	framerule=0.5pt,
	xleftmargin=0.6em,
	xrightmargin=0.6em,
	framesep=6pt,
	emph={[1]User},
	emphstyle=[1]\color{conversationUser}\bfseries,
	emph={[2]System},
	emphstyle=[2]\color{conversationSystem}\bfseries,
	emph={[3]VenueAccessibility,DietaryPolicy,InvoiceRequest,Resume,EventRegistration},
	emphstyle=[3]\color{conversationNote}
}

\tikzset{
	godr_box/.style={
		draw=lineDark,
		rounded corners=3pt,
		very thick,
		align=center,
		font=\sffamily\small,
		text=textDark,
		minimum height=0.82cm,
		inner xsep=7pt,
		inner ysep=5pt
	},
	godr_core/.style={godr_box, fill=layerGoal, minimum width=6.6cm},
	godr_sub/.style={godr_box, fill=white, minimum width=2.55cm, font=\sffamily\scriptsize},
	runtime_box/.style={godr_box, fill=layerExec, minimum width=6.6cm},
	layer_box/.style={godr_box, minimum width=4.7cm, minimum height=1.05cm},
	workflow_node/.style={godr_box, fill=white, minimum width=3.8cm, minimum height=0.72cm, font=\sffamily\scriptsize},
	workflow_root/.style={godr_box, fill=layerExec, minimum width=3.8cm, minimum height=0.78cm, font=\sffamily\small\bfseries},
	policy_note/.style={godr_box, fill=layerPolicy!45, minimum width=3.7cm, align=left, font=\sffamily\scriptsize},
	godr_arrow/.style={-{Latex[length=2.5mm]}, very thick, draw=lineDark},
	workflow_arrow/.style={-{Latex[length=2.0mm]}, thick, draw=lineDark!75},
	trace_arrow/.style={-{Latex[length=2.0mm]}, thick, draw=lineDark!70, dashed}
}

\title{From Task-Guided Conversational Graphs to Goal-Oriented Dialogue Runtimes}

\newif\ifuniqueAffiliation
\uniqueAffiliationtrue

\ifuniqueAffiliation 
\author{ \href{https://orcid.org/0009-0008-0201-2984}{\includegraphics[scale=0.06]{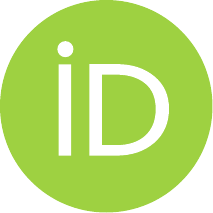}\hspace{1mm}Mariano Garralda-Barrio}\thanks{Independent Researcher / Investigador Independiente.} \\
Independent Researcher\\
Lleida, Spain \\
\texttt{mariano.garralda.r@gmail.com} \\
}
\fi

\renewcommand{\headeright}{}
\renewcommand{\undertitle}{A Conceptual Framework for Conversational Architecture Engineering}

\begin{document}

\maketitle

\begin{abstract}
Graph and multi-agent orchestration frameworks make production large language model (LLM) workflows practical, but they do not by themselves solve conversational continuity when users maintain several interdependent objectives. This conceptual systems paper focuses on the high-complexity end of that design space, where goals can be suspended, resumed, revised, and invalidated by actions in other goals. We introduce the Goal-Oriented Dialogue Runtime (GODR), a framework-neutral design pattern that treats goals, task frames, lifecycle state, invalidation rules, and resumption contracts as first-class runtime objects while delegating bounded execution to graph runtimes, agents, tools, or application programming interfaces (APIs). GODR is not proposed as a replacement for workflow graphs in simple guided processes; it is intended for complex, multi-domain, interruptible conversations where objective continuity cannot be recovered reliably from agent identity, chat history, or execution-graph position alone. The paper formalizes the problem, proposes runtime objects and architecture-selection criteria, and frames evaluation as an agenda for future empirical validation rather than as a measured performance claim.
\end{abstract}

\section{Introduction}

Conversational artificial intelligence (AI) engineering is increasingly moving from prompt-centric interactions toward stateful, tool-augmented, and agent-oriented systems~\cite{yao2023react,shu2022dialog2api}. Current orchestration frameworks expose agents, tools, and multi-agent orchestration patterns as practical execution primitives \citep{langchain_multiagent,google_adk_patterns,microsoft_agent_framework,openai_agents}. They also provide mechanisms for subgraphs, handoffs, memory, checkpoints, and tracing \citep{langgraph_subgraphs,langchain_handoffs,google_adk_memory}. These primitives are valuable for execution continuity, but they do not fully solve conversations where several user objectives remain active, share constraints, and can invalidate one another.

This paper proposes the Goal-Oriented Dialogue Runtime (GODR) as a framework-neutral layer for making those objectives explicit. More broadly, GODR follows the same architectural direction as recent agent-system work that treats runtime state, tools, memory, orchestration, and evaluation as explicit engineering objects rather than as prompt-only concerns~\cite{xu2026agents,garraldabarrio2026governedevolutionagentruntimes}.

In a \emph{process-driven conversation}, the system guides the user through a known sequence: identify the customer, collect data, validate preconditions, request approval, and close the operation. This is the type of architecture where a root graph with business subgraphs works well. The user may retry an answer, trigger a human-in-the-loop step, cancel the process, or reach a reset condition, but the conversational freedom remains bounded by the process.

The harder case follows a different organizing principle. Rather than being driven by a predefined process, the conversation is organized around a user objective that remains active across interruptions, subgoals, and contextual shifts. We refer to this as a \emph{conversation-driven goal}: a conversational objective whose continuity cannot be reduced to a single execution path. For example, a user registering for a professional workshop may branch into venue accessibility, dietary requirements, hotel logistics, session selection, payment constraints, group discounts, invoice requests, or external interruptions. Some branches are side questions; others become subgoals; others supersede the original objective. The system must decide not only which agent answers next, but which user goal remains alive, which goal is suspended, and how to resume it.

The central claim is deliberately scoped: conventional finite-state machines (FSMs), workflow graphs, or goal stacks are often sufficient for bounded processes and shallow interruptions~\cite{bohus2009ravenclaw,microsoft_waterfall}. Full GODR becomes useful when the conversation contains multiple open goals, non-local dependencies, shared constraints, and invalidation events. In that regime, agent graphs and workflow graphs remain necessary execution substrates, but the runtime also needs an explicit object for objective ownership. GODR treats execution as a service of goals rather than treating goals as incidental attributes of execution.

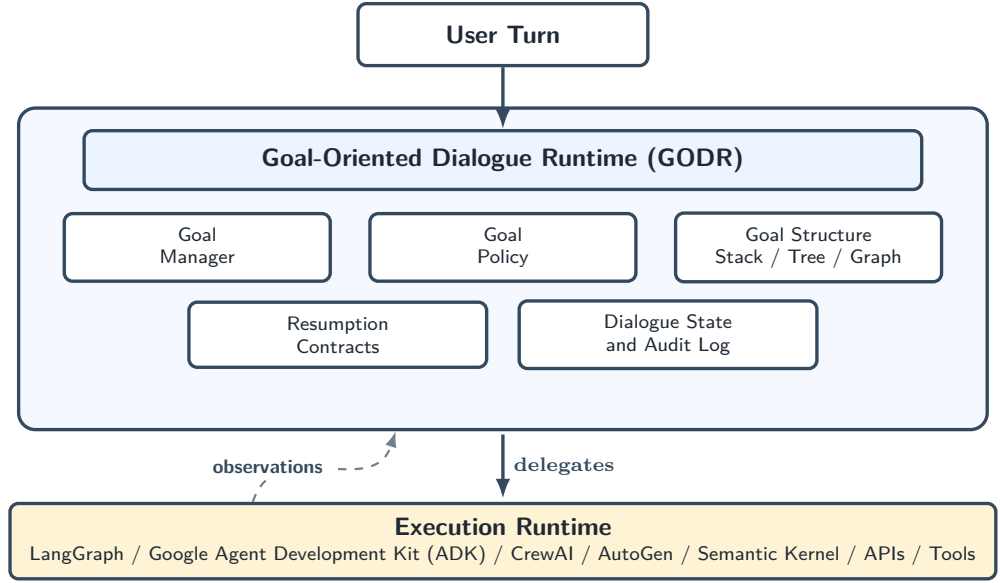
\begin{figure}[t]
\centering
\resizebox{0.80\linewidth}{!}{%
\begin{tikzpicture}[
	every node/.style={font=\sffamily},
	architecture_panel/.style={draw=lineDark, rounded corners=6pt, very thick, fill=layerGoal!28, minimum width=12.2cm, minimum height=4.05cm},
	architecture_header/.style={godr_box, fill=layerGoal!55, minimum width=10.55cm, minimum height=0.74cm, font=\sffamily\small\bfseries},
	architecture_card/.style={godr_box, fill=white, minimum width=3.35cm, minimum height=0.82cm, font=\sffamily\scriptsize},
	architecture_wide_card/.style={godr_box, fill=white, minimum width=3.75cm, minimum height=0.82cm, font=\sffamily\scriptsize},
	feedback_label/.style={fill=white, inner xsep=4pt, inner ysep=2pt, font=\sffamily\scriptsize\bfseries, text=lineDark}
]

\node[godr_box, fill=white, minimum width=3.65cm, minimum height=0.78cm] (user) at (0,0) {\textbf{User Turn}};

\node[architecture_panel] (godrpanel) at (0,-2.95) {};
\node[architecture_header] (godr) at (0,-1.58) {Goal-Oriented Dialogue Runtime (GODR)};

\node[architecture_card] (manager) at (-3.85,-2.68) {Goal\\Manager};
\node[architecture_card] (policy) at (0,-2.68) {Goal\\Policy};
\node[architecture_card] (structure) at (3.85,-2.68) {Goal Structure\\Stack / Tree / Graph};
\node[architecture_wide_card] (resume) at (-2.08,-3.78) {Resumption\\Contracts};
\node[architecture_wide_card] (state) at (2.08,-3.78) {Dialogue State\\and Audit Log};

\node[runtime_box, minimum width=10.8cm, minimum height=0.92cm] (exec) at (0,-6.38) {\textbf{Execution Runtime}\\[-1pt]{\scriptsize LangGraph / Google Agent Development Kit (ADK) / CrewAI / AutoGen / Semantic Kernel / APIs / Tools}};

\draw[godr_arrow] (user) -- (godr);
\draw[godr_arrow] ([yshift=-0.04cm]godrpanel.south) -- node[right, pos=0.50, font=\scriptsize\bfseries, text=lineDark] {delegates} ([yshift=0.04cm]exec.north);
\draw[trace_arrow] ([xshift=-3.15cm]exec.north) to[out=72,in=-108,looseness=1.18]
	node[pos=0.56, left, feedback_label] {observations} ([xshift=-1.35cm]godrpanel.south);

\end{tikzpicture}
}
\caption{
Proposed Goal-Oriented Dialogue Runtime within the conversational stack. The figure illustrates the central hypothesis of this paper: goal management should be represented as an explicit runtime layer above execution frameworks rather than being implicitly encoded in agents, memory structures, or workflow graphs.}
\label{fig:godr-architecture}
\end{figure}

Figure~\ref{fig:godr-architecture} summarizes the core architecture: goals are managed above execution frameworks, not hidden inside them.

This paper makes four contributions:

\begin{itemize}[leftmargin=*]
\item It defines the \emph{Multi-Objective Interruptible Dialogue Problem}, separating active-agent selection and execution continuity from active-goal continuity.
\item It introduces \emph{goal complexity} as a taxonomy for selecting conversational architectures and relating stacks, trees, and goal graphs to underlying dependency structure.
\item It proposes the \emph{Goal-Oriented Dialogue Runtime} as a framework-neutral layer above graph and agent execution substrates, with goals, lifecycles, task frames, invalidation rules, and resumption contracts as first-class runtime objects.
\item It provides an engineering methodology, evaluation criteria, and baselines for selecting and testing dialogue architectures under interruptible multi-domain goal complexity.
\end{itemize}

Operationally, GODR is a runtime dedicated to objective continuity. It maintains goal state, lifecycle policies, resumption contracts, and inter-goal consistency rules while delegating bounded execution to graph runtimes, tools, and specialist agents. The design boundary is explicit: execution frameworks manage how work is performed; GODR manages why the work remains relevant and how interrupted objectives are resumed.

\section{Motivating Example: From Warranty Claims to Event Registration}

The contrast between process-driven and conversation-driven systems is easiest to see through two service scenarios. Both can use agents, tools, and graph runtimes, but they differ in what must remain stable across turns: the process position in the first case, and the user's suspended objective in the second.

\subsection{Process-Driven Conversational Workflow}

A warranty replacement assistant is a representative example of a process-driven conversation. The user can provide information in natural language, but the system still owns the sequence: identify the account, validate the warranty, diagnose the issue, approve replacement, and collect shipping details.

\begin{lstlisting}[style=conversation]
User: My noise-cancelling headphones stopped charging.
System: I can help with a warranty replacement. What is the order number?
User: It is ORD-3147.
System: The device is still under warranty. Did the charging indicator light turn on
        when you tried a different cable?
User: No, I tried two cables and nothing lights up.
System: The issue qualifies for replacement. Please confirm the shipping
        address for the replacement unit.
\end{lstlisting}

This dialogue can be implemented naturally as a root graph with business subgraphs and global human-in-the-loop or reset policies.

The architecture remains understandable because the main object being executed is a process. The user may answer freely, retry, cancel, or escalate, but the conversational path remains anchored in the warranty workflow.

\subsection{Conversation-Driven Task Orchestration}

An event registration assistant has a different shape. The user starts with a registration goal, but the conversation can legitimately branch into venue, catering, billing, travel, or policy questions before returning to the unfinished registration.

\begin{lstlisting}[style=conversation]
User: I want to register for the AI Systems Workshop next Friday.
System: I found two sessions, morning and afternoon. Which do you prefer?
User: Before choosing, is the venue wheelchair accessible?
System: Yes. The venue has step-free access and accessible seating.
        [VenueAccessibility goal completed]
User: Good. Can I request a vegan meal?
System: Yes. I can add a vegan meal request to the registration.
        [DietaryPolicy goal completed]
User: Also, will you send an invoice to my company?
System: Yes. I can collect billing details after the session is selected.
        [InvoiceRequest goal suspended]
User: Continue with the registration.
System: Resuming registration: morning or afternoon session?
        [Resume EventRegistration at waiting_for_session_preference]
\end{lstlisting}

A fully connected graph of agents can route these turns, but routing is not enough. The unresolved object is the interrupted registration goal. The system must preserve the selected event, the missing session preference, the side constraints introduced by accessibility and catering, and the exact resumption point.

\section{Execution Continuity versus Objective Continuity}
\begin{figure}[t]
\centering
\resizebox{0.85\linewidth}{!}{%
\begin{tikzpicture}[node distance=1.55cm, every node/.style={font=\sffamily}]
\node[workflow_root] (current) {current\_node};
\node[workflow_node, below left=of current] (retry) {retry(current\_node)};
\node[workflow_node, below=of current] (continue) {continue(next\_node)};
\node[workflow_node, below right=of current] (reset) {abort\_or\_reset(process)};

\draw[workflow_arrow] (current) -- (retry);
\draw[workflow_arrow] (current) -- (continue);
\draw[workflow_arrow] (current) -- (reset);
\end{tikzpicture}
}
\caption{Local control moves in a process-guided dialogue. The runtime retries the current node, advances to the next node, or aborts and resets the process.}
\label{fig:process-control-moves}
\end{figure}
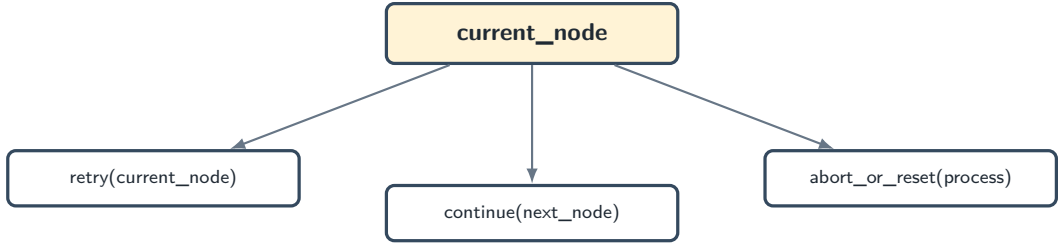

The boundary between process-guided and multi-goal dialogue systems can also be described in terms of control continuity. In a process-guided conversational system, continuity is mostly equivalent to preserving the current process position. From a given execution node, the runtime usually needs only three control moves: retry the same node, advance to the next node, or abort or reset the process. Figure~\ref{fig:process-control-moves} summarizes this local control model.

This pattern is not a limitation of workflow-based systems; it is precisely where they are strongest. Warranty replacement, onboarding, travel booking, know-your-customer checks, claims handling, and similar enterprise workflows often have a single active trajectory. Local retries, validation loops, escalation policies, and global resets can be represented cleanly because the current execution position and the user's conversational objective remain largely aligned.

Importantly, the complexity of a workflow node is orthogonal to the complexity of the goal structure. A node may internally contain a ReAct agent~\cite{yao2023react}, planning loops, tool orchestration, or human-in-the-loop interactions while still preserving a single execution position and a single active objective. Such mechanisms increase local reasoning and execution complexity, but do not by themselves require objective-level lifecycle management.

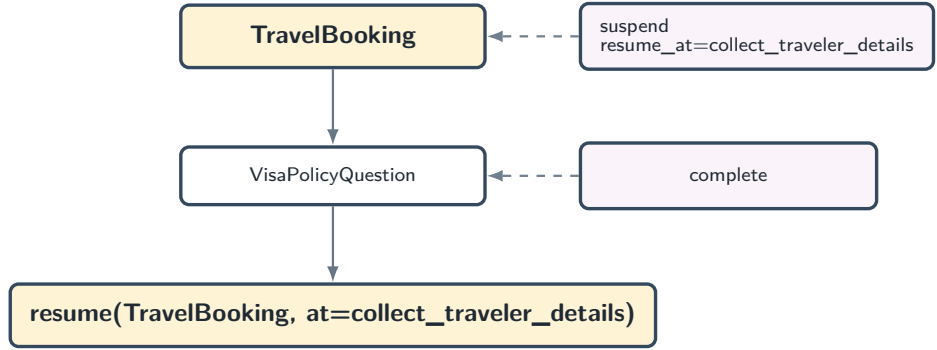
\begin{figure}[t]
\centering
\resizebox{0.75\linewidth}{!}{%
\begin{tikzpicture}[node distance=0.95cm, every node/.style={font=\sffamily}]
\node[workflow_root] (travel) {TravelBooking};
\node[policy_note, right=1.15cm of travel] (suspend) {suspend\\[-1pt]{\scriptsize resume\_at=collect\_traveler\_details}};
\node[workflow_node, below=of travel] (policy) {VisaPolicyQuestion};
\node[policy_note, right=1.15cm of policy] (complete) {complete};
\node[workflow_root, below=of policy, minimum width=5.4cm] (resume) {resume(TravelBooking, at=collect\_traveler\_details)};

\draw[workflow_arrow] (travel) -- (policy);
\draw[workflow_arrow] (policy) -- (resume);
\draw[trace_arrow] (suspend) -- (travel.east);
\draw[trace_arrow] (complete) -- (policy.east);
\end{tikzpicture}
}
\caption{Objective continuity across an interruption. The travel booking remains resumable while the visa-policy question temporarily controls the dialogue.}
\label{fig:objective-continuity}
\end{figure}

The conceptual break appears when user goals become suspendible. A side question, policy check, invoice request, or dependent subtask may temporarily take control while the original objective remains alive. The system must then preserve not only the current execution node, but also the suspended goal, its pending action, the logical return point, the context required for safe resumption, and any invalidation conditions introduced by later turns. Figure~\ref{fig:objective-continuity} illustrates this distinction with a travel-booking goal interrupted by a visa-policy question.

In this regime, the runtime no longer only advances, retries, or restarts. It performs a logical return to a suspended objective without rolling back execution history. This is the role of a resumption contract: it captures the continuation that must survive interruptions, tool calls, agent changes, and later user corrections. Process-guided systems primarily require execution continuity, whereas multi-goal conversational systems require objective continuity.

\section{Goal-Complexity Taxonomy}

Conversational type is not enough; designers also need to estimate the complexity of the goal structure itself. Table~\ref{tab:goal-complexity} defines five goal-complexity (GC) levels used throughout the paper. When graph structure is needed, a directed acyclic graph (DAG) is a useful representation for dependencies without cycles:

\begin{table}[t]
	\centering
	\caption{Goal complexity levels for conversational runtime selection.}
	\label{tab:goal-complexity}
	\small
	\begin{tabularx}{\linewidth}{p{0.11\linewidth}Y Y p{0.18\linewidth}}
	\toprule
	\textbf{Level} & \textbf{Goal structure} & \textbf{Example} & \textbf{Suggested model} \\
	\midrule

	GC-0 &
	Single immutable goal following a fixed execution path. &
	Interactive voice response troubleshooting. &
	FSM \\

	GC-1 &
	Single goal with bounded interruptions, retries, or escalation paths. &
	Warranty assistant with escalation. &
	Workflow graph \\

	GC-2 &
	One active goal path at a time, with nested goals and explicit resumption points. &
	Event registration with side questions. &
	Goal stack \\

	GC-3 &
	Multiple independent goals remain active simultaneously and compete for attention. &
	Enterprise copilot tracking ongoing user tasks. &
	Goal tree or agenda \\

	GC-4 &
	Multiple active goals with dependency, coordination, and invalidation relationships. &
	Operational AI assistant coordinating approvals, resources, and policies. &
	GODR + goal graph / DAG \\

	\bottomrule
	\end{tabularx}
\end{table}

The key design threshold for this paper is GC-4. GC-0 and GC-1 are usually well served by FSMs or workflow graphs; GC-2 often needs a goal stack; GC-3 may require an agenda or goal tree. Below GC-4, the cost of explicit goal lifecycle management often exceeds its practical benefit. GODR is intended for GC-4, where several goals remain alive and actions in one goal can change the validity, priority, or resumability of another. This is where dependency and invalidation semantics become runtime concerns rather than implementation details.

\subsection{A GC-4 Example: Corporate Procurement Assistant}

A corporate procurement assistant illustrates why a goal graph is sometimes necessary. Consider a session in which a user asks the assistant to purchase hardware for a new team. The session may contain several open goals, with dependencies and invalidation relations like those in Figure~\ref{fig:procurement-goal-graph}:

\begin{figure}[H]
\centering
\resizebox{0.75\linewidth}{!}{%
\begin{tikzpicture}[
	every node/.style={font=\sffamily\scriptsize},
	goal_main/.style={workflow_root, fill=layerGoal, minimum width=4.2cm},
	goal_dep/.style={workflow_node, minimum width=3.45cm},
	goal_leaf/.style={workflow_node, minimum width=3.35cm},
	change_event/.style={policy_note, fill=layerPolicy!50, minimum width=3.65cm, align=center},
	relation_arrow/.style={-{Latex[length=2.0mm]}, thick, draw=lineDark!68},
	invalidation_arrow/.style={-{Latex[length=2.0mm]}, thick, draw=lineDark!68, dashed}
]
\node[goal_main] (purchase) at (0,4.55) {PurchaseRequest};
\node[goal_dep] (vendor) at (0,3.05) {VendorSelection};
\node[goal_leaf] (budget) at (-5.30,1.55) {BudgetApproval};
\node[goal_leaf] (compliance) at (-1.35,1.55) {ComplianceCheck};
\node[goal_leaf] (delivery) at (3.20,1.55) {DeliveryConstraint};
\node[change_event] (change) at (-4.95,-0.35) {VendorChangeEvent\\[-1pt]{\scriptsize triggers invalidation}};
\node[goal_dep] (invoice) at (0.35,-0.35) {InvoiceGeneration};

\draw[relation_arrow] (purchase.south) -- (vendor.north);
\draw[relation_arrow] (purchase.south west) -- (budget.north);
\draw[relation_arrow] (vendor.south) -- (compliance.north);
\draw[relation_arrow] (vendor.south east) -- (delivery.north);
\draw[relation_arrow] (budget.south) to[out=-30,in=160] (invoice.north west);
\draw[relation_arrow] (compliance.south) -- (invoice.north);
\draw[relation_arrow] (delivery.south) -- (invoice.north east);
\draw[invalidation_arrow] (change.east) -- (invoice.west);
\draw[invalidation_arrow] (change.north east) to[out=45,in=-120] (compliance.south west);
\end{tikzpicture}
}
\caption{A GC-4 procurement goal graph. Solid arrows encode subgoal and required-for relations; dashed arrows encode invalidation events. Invoice generation depends on budget approval, compliance, and delivery constraints; a vendor-change event invalidates both compliance and invoice generation.}
\label{fig:procurement-goal-graph}
\end{figure}
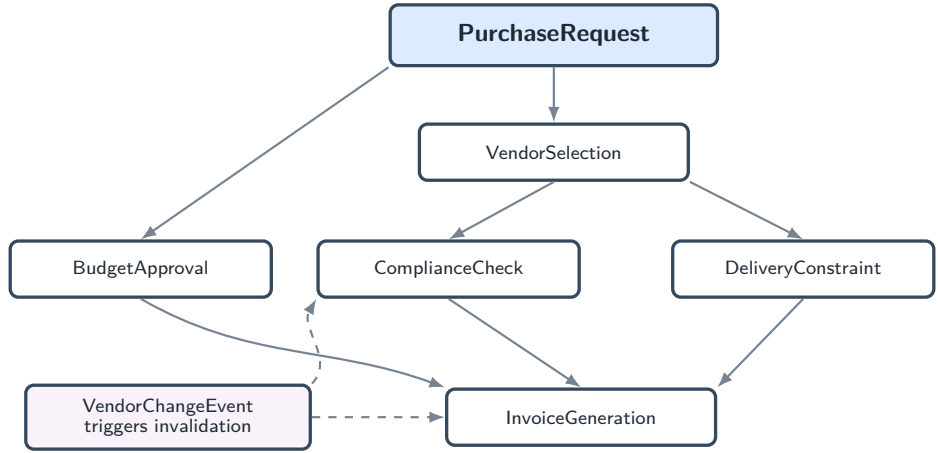

This is not a stack. The user may first request a purchase, then ask for vendor alternatives, then trigger a compliance check, then revise the budget, and later ask for invoice preparation. If the budget is rejected, the purchase request is blocked; if the vendor changes, the compliance check and invoice draft may be invalidated; if delivery constraints change, vendor selection may need to be reopened without cancelling the original purchase request.

A root graph can encode one procurement process, but the conversational session may contain multiple partially complete, cross-dependent goals. A supervisor can route between procurement, compliance, budget, and invoicing agents, but routing alone does not express which goals are invalidated by a vendor change or which suspended goals remain safe to resume. This is the GC-4 regime: the runtime requires a goal DAG with explicit dependency, blocking, supersession, and invalidation relations.

\section{Related Work}

The proposal builds on several research and engineering traditions rather than starting from a blank slate. This section reviews the most relevant precedents: classic task-oriented dialogue systems, multi-domain dialogue state tracking, planning and Belief--Desire--Intention (BDI) models, and recent LLM orchestration frameworks. The goal is to clarify which ideas GODR reuses and where it introduces a distinct runtime boundary for conversational goal continuity.

The evolution of conversational architectures can be interpreted as a progressive externalization of conversational state and control. Early dialogue systems externalized dialogue state through information-state and dialogue-management models. Multi-domain systems extended this idea to goal, schema, and service tracking. Planning and BDI architectures introduced explicit representations of objectives, commitments, and action structure. Modern LLM frameworks externalize execution through agents, workflows, tools, and orchestration runtimes. GODR follows this trajectory by externalizing conversational goal lifecycle management as a runtime concern.

\subsection{Classic Task-Oriented Dialogue Systems}

Pre-LLM task-oriented dialogue systems did not rely primarily on a single giant graph. The canonical architecture separated natural language understanding, dialogue state tracking, dialogue management, policy, and response generation, often through explicit information-state or decision-theoretic dialogue models \citep{larsson2000informationstate,williams2007pomdp,young2013pomdp}. RavenClaw is particularly relevant because it models dialogue management through hierarchical task decomposition and an expectation agenda \citep{bohus2003ravenclaw,bohus2009ravenclaw}. Its agenda-based interpretation anticipates the need to represent pending expectations rather than only active transitions.

Microsoft Bot Framework formalized the Dialog Stack: a dialog context contains active dialogs, and the dialog stack acts as a call stack for them \citep{microsoft_dialogs,microsoft_waterfall}. This is a concrete engineering precedent for suspended and resumed subdialogues.

\subsection{Multi-Domain Dialogue State Tracking}

Multi-domain dialogue state tracking has long been treated as a central problem for task-oriented assistants, from shared tracking challenges and large multi-domain corpora to schema-guided service representations \citep{henderson2014dstc,budzianowski2018multiwoz,rastogi2020sgd}. The Schema-Guided Dialogue dataset highlights the scalability problem in multi-domain assistants. It provides annotations for intent prediction, slot filling, dialogue state tracking, and response generation across many domains and services \citep{rastogi2020sgd,google_sgd_blog}. Importantly, it frames state tracking as estimating the user goal over dynamic service schemas.

Recent work such as Dialog2API makes the connection clearer: it represents dialogue state as a stack of programs, with the most recently mentioned program at the top \citep{shu2022dialog2api}. This is conceptually close to a goal stack or task stack for composite, revisable user objectives. Task-Oriented Dialogue as Dataflow Synthesis takes another route by representing dialogue state as a dataflow graph \citep{andreas2020dataflow}. The common theme is that mature dialogue systems externalize state and task structure instead of encoding the whole interaction as a flat transition graph.

\subsection{Relationship with Classical Planning}

The proposed model is related to classical planning, but it addresses a different runtime problem. Planning typically reasons over actions, preconditions, effects, and goals to synthesize or select a sequence of actions that reaches a desired state \citep{ghallab2004automatedplanning}. GODR is concerned with the lifecycle of user goals during an interactive session: goals may be partially specified, suspended, resumed, revised, abandoned, superseded, or invalidated by later dialogue.

The distinction is operational. A planner may decide how to satisfy a booking objective; a goal-oriented dialogue runtime decides whether the booking objective is still active, whether it has been suspended by a side question, whether its resumption contract remains valid, and whether later user input has revised or invalidated it. GODR can therefore use planning inside the execution layer, but it should not be reduced to planning alone.

\subsection{Goal Decomposition and HTN Planning}

Hierarchical Task Network (HTN) planning decomposes high-level tasks into structured networks of subtasks and ordering constraints, and it has long served as a model for goal-directed decomposition and execution control \citep{erol1994htn,erol1996htn,ghallab2004automatedplanning}. GODR shares the idea that complex objectives may require subordinate structure: a conversational goal can spawn subgoals, maintain pending actions, or depend on other goals.

The boundary is different, however. HTN planning primarily focuses on plan construction, task decomposition, and execution ordering. GODR focuses on conversational goal lifecycle management: interruption handling, suspension, resumption, supersession, cross-goal invalidation, and auditability during dialogue. In this sense, HTN planning can be viewed as a potential execution strategy within a goal, whereas GODR governs the persistence and coordination of goals themselves.

\subsection{Relationship with BDI Architectures}

The terminology of goals, policies, and intentions naturally recalls Belief--Desire--Intention (BDI) architectures, where agents maintain informational beliefs, motivational desires, and committed intentions \citep{bratman1987intention,rao1995bdi}. GODR is compatible with that lineage but focuses on a narrower systems problem: preserving conversational goal continuity across modern LLM orchestration substrates.

In BDI terms, a suspended goal resembles an intention that should not be forgotten merely because another intention temporarily controls behavior. However, GODR does not require a full BDI agent model. Its contribution is architectural: it externalizes goal lifecycle, goal structure, resumption contracts, and invalidation relations as runtime objects that can sit above graph runtimes, tool agents, or workflow engines.

\subsection{Modern LLM Orchestration Frameworks}

Modern orchestration frameworks such as LangGraph, Google Agent Development Kit (ADK), Semantic Kernel and Microsoft Agent Framework, CrewAI, OpenAI Agents, Amazon Bedrock Agents, and AutoGen provide various combinations of agents, workflows, tools, handoffs, state management, memory, checkpoints, tracing, and multi-agent coordination \citep{langgraph_subgraphs,langchain_handoffs,google_adk_patterns,google_adk_memory,microsoft_semantic_kernel_agents,microsoft_semantic_kernel_orchestration,microsoft_agent_framework,crewai_flows,crewai_flow_state,openai_agents,aws_bedrock_multiagent,autogen_multiagent}.

These systems differ substantially in programming model and operational scope, but they support the same broad architectural trend: execution is externalized into explicit runtimes, graphs, agents, tools, and workflow substrates. That is the layer GODR assumes rather than replaces.

Across these frameworks, the common primitives are agents, tools, workflows, handoffs, state, and memory. Table~\ref{tab:framework-comparison} therefore compares the execution responsibilities commonly provided by modern orchestration frameworks with the objective-continuity responsibilities introduced by GODR.

\begin{table}[t]
	\centering
	\caption{Execution continuity provided by modern orchestration frameworks versus objective-continuity responsibilities introduced by GODR.}
	\label{tab:framework-comparison}
	\scriptsize
	\begin{tabularx}{\linewidth}{p{0.30\linewidth}Y Y}
	\toprule
	\textbf{Capability} & \textbf{Modern orchestration frameworks} & \textbf{GODR layer} \\
	\midrule
	Agent and workflow orchestration & First-class agents, graphs, handoffs, supervisors, or flows. & Delegated to external runtimes. \\
	Tool and API execution & First-class tool invocation and bounded task execution. & Delegated through runtime adapters. \\
	State, memory, and checkpoints & Supported through framework-specific state, memory, tracing, or persistence mechanisms. & Reused as execution evidence, rather than the authoritative goal state. \\
	Goal lifecycle & Typically application-defined above the framework state model. & First-class runtime concern: \texttt{created}, \texttt{active}, \texttt{suspended}, \texttt{blocked}, and terminal states. \\
	Resumption contracts & Typically application-defined through checkpoints, prompts, or custom state. & First-class contract with pending action, required context, return point, and invalidation conditions. \\
	Goal structures & Typically application-defined as stacks, task trees, graph conventions, or ad hoc registries. & First-class runtime representation of conversational goal structures. \\
	Cross-goal semantics & Typically application-defined through business rules or controller code. & Runtime-managed dependency, invalidation, blocking, supersession, and consistency relations. \\
	Objective continuity & Emerges from application logic built on top of execution state. & Explicit responsibility of the runtime layer. \\
	\bottomrule
	\end{tabularx}
\end{table}

This comparison is not a claim that existing frameworks are incomplete for their intended purpose. Rather, it identifies a missing layer: they provide strong execution and orchestration substrates, while goal lifecycle and objective continuity remain design responsibilities for the application architect.

Taken together, these lines of work reveal a recurring pattern. Dialogue systems externalize state; planning systems externalize action selection and task decomposition; BDI systems externalize intentions; and modern orchestration frameworks externalize execution. However, conversational goal lifecycle management remains largely embedded in application-specific logic. GODR is proposed as an explicit runtime abstraction for this remaining concern.

\section{Problem Statement}

Current LLM orchestration frameworks are effective at answering: which node, agent, or tool should run next? The harder question in multi-objective conversations is: which user goal is active, which goals are suspended, and what is the correct resumption point?

We define the \emph{Multi-Objective Interruptible Dialogue Problem} as follows: given a conversation history $H$, a set of active and suspended goals $G$, a set of available agents and tools $A$, and a new user utterance $u_t$, determine which goal-level operation in Table~\ref{tab:decision-space} should update the active goal structure.

\begin{table}[t]
\centering
\caption{Goal-policy decision space at turn $t$.}
\label{tab:decision-space}
\small
\begin{tabular}{p{0.28\linewidth}p{0.60\linewidth}}
\toprule
\textbf{Operation} & \textbf{Runtime meaning} \\
\midrule
\texttt{continue(current\_goal)} & Advance the active goal with the new user contribution. \\
\texttt{revise(current\_goal)} & Update constraints, slots, or assumptions for the active goal. \\
\texttt{push(child\_goal)} & Create a nested goal owned by the current goal. \\
\texttt{switch(goal\_i)} & Transfer control to another open goal. \\
\texttt{pop(completed\_goal)} & Close a completed goal and return to the appropriate parent or agenda item. \\
\texttt{resume(previous\_goal)} & Reactivate a suspended goal at its resumption contract. \\
\texttt{cancel(goal\_i | all\_goals)} & Abandon one goal or the active goal structure. \\
\texttt{escalate\_to\_human(goal\_i)} & Transfer a goal to a human decision or review path. \\
\texttt{reset\_on\_failure} & Reset a failed goal or the entire session under policy. \\
\bottomrule
\end{tabular}
\end{table}

This decision cannot be solved cleanly by agent routing alone because the active agent and the active goal are not equivalent. A venue-information agent may answer a side question inside an event-registration goal, or it may own an independent venue-information goal. This semantic distinction matters for memory, resumption, user experience, and auditability.

Not all interruptions imply a goal transition. Some interruptions correspond to associated objectives that semantically support completion of the current goal, such as requesting clarification about a required document, a policy, or a business rule. In these cases, conversational control may temporarily shift, but the parent goal remains the primary objective driving the interaction. Other interruptions introduce independent goals with their own completion criteria and lifecycle. Distinguishing associated from independent goals is therefore an implementation concern of the Goal Policy rather than a separate goal-complexity category: both may appear within the same GC level, but they may trigger different runtime operations such as \texttt{continue}, \texttt{revise}, \texttt{push}, \texttt{switch}, or \texttt{resume}.

\section{Operational Model}

A Goal-Oriented Dialogue Runtime can be described as a transition system over conversational goals. At turn $t$, the dialogue state is represented as

\begin{equation}
D_t = \langle H_t, G_t, a_t, C_t, L_t \rangle,
\end{equation}

where $H_t$ is the conversation history, $G_t$ is the current goal structure, $a_t$ is the active goal identifier, $C_t$ is the set of global session constraints, and $L_t$ is the audit log. The goal structure is a labeled directed graph

\begin{equation}
G_t = (V_t, E_t, \lambda_V, \lambda_E),
\end{equation}

where each $v \in V_t$ is a goal object, each $e \in E_t$ is a relation between goals, $\lambda_V$ assigns lifecycle and frame attributes to goals, and $\lambda_E$ assigns relation types such as \texttt{parent}, \texttt{depends\_on}, \texttt{blocks}, \texttt{supersedes}, \texttt{resumes}, or \texttt{invalidates}. A stack is the special case where $G_t$ is a linear chain with last-in-first-out control. A tree is the special case where relations are hierarchical. A DAG is required when dependency or invalidation relations cross hierarchical branches.

A goal is a persistent conversational objective whose lifecycle may extend across multiple turns, agents, tools, and execution graphs, and whose completion or validity cannot be inferred solely from local execution state. This definition distinguishes goals from turn-level intents and from local workflow progress: a goal is the object that remains accountable for continuity when the conversation is interrupted, revised, resumed, or invalidated.

A goal $g \in V_t$ is represented as

\begin{equation}
g = \langle id, type, status, frame, agent, graph, resume, policy \rangle,
\end{equation}

where \textit{frame} stores known values, missing slots, constraints, and local memory; \textit{resume} is a resumption contract; and \textit{policy} encodes local safety or business constraints.

This definition also separates GODR from nearby abstractions that are often conflated in LLM systems. An intent classifies the user's current utterance; a memory stores reusable information; a plan proposes future actions; and workflow state records progress inside a bounded execution process. GODR governs the lifecycle of user objectives across these objects: it decides which objective is active, suspended, resumed, invalidated, or closed. Table~\ref{tab:goal-boundary} makes this boundary explicit. In this paper, an object belongs to GODR only when it participates in goal lifecycle transitions such as activation, suspension, resumption, revision, invalidation, supersession, completion, cancellation, or escalation.

\begin{table}[t]
\centering
\caption{Boundary between goal-runtime objects and adjacent conversational abstractions.}
\label{tab:goal-boundary}
\small
\begin{tabular}{p{0.22\linewidth}p{0.33\linewidth}p{0.34\linewidth}}
\toprule
\textbf{Abstraction} & \textbf{Primary role} & \textbf{Why it is not enough for goal continuity} \\
\midrule
Intent & Classifies what the latest user turn appears to request. & It is turn-local and does not encode suspended objectives or resumption points. \\
Memory & Stores reusable facts, preferences, or prior conversation content. & It represents information but does not govern lifecycle transitions. \\
Plan & Orders proposed actions for a task or agent. & It may change without preserving which user objective owns the plan. \\
Workflow state & Tracks progress inside a bounded business or tool process. & It is local to execution and does not manage cross-goal interruption or invalidation. \\
Task frame & Stores slots, constraints, and local state for a goal. & It needs an owning goal and lifecycle policy to remain resumable and auditable. \\
\bottomrule
\end{tabular}
\end{table}

Given a user utterance $u_t$, the Goal Policy computes a goal operation

\begin{equation}
o_t = \pi(u_t, D_t) \in \mathcal{O},
\end{equation}

where $\mathcal{O}=\{continue, revise, push, switch, pop, resume, cancel, escalate, reset\}$. The operation transforms the dialogue state before the selected graph runtime or agent is invoked:

\begin{equation}
D_{t+1}^{pre} = T_o(D_t, o_t), \qquad y_t = E(D_{t+1}^{pre}, u_t), \qquad D_{t+1}=U(D_{t+1}^{pre}, y_t).
\end{equation}

Here $T_o$ is the goal-level transition, $E$ is the bounded execution substrate, and $U$ is the state update induced by execution results. This separation is central: graph runtimes execute local task logic, while the Dialogue Manager preserves goal continuity across interruptions.

\subsection{Goal-Structure Invariants}

The model is useful only if the goal structure obeys explicit invariants. A production implementation should enforce at least the following:

\begin{enumerate}[leftmargin=*]
\item \textbf{Unique active goal.} At most one goal has status \texttt{active} for a session unless the system explicitly enters a parallel-goal mode.
\item \textbf{Resumability of suspended goals.} Every suspended goal must have a non-empty resumption contract or be marked \texttt{blocked}, \texttt{abandoned}, \texttt{superseded}, or \texttt{failed}.
\item \textbf{Structure-registry consistency.} Every identifier in the stack, tree, or graph must exist in the goal registry and refer to a non-terminal goal.
\item \textbf{Frame ownership.} Updates to task-frame fields must be attributed to a goal, agent, tool, or human decision.
\item \textbf{Invalidation safety.} Resuming a goal requires checking its invalidation conditions against newer dialogue events and global constraints.
\item \textbf{Auditable transitions.} Every operation that changes goal status, stack order, or resumption contract must be logged as a goal-level event.
\end{enumerate}

These invariants make the proposed abstraction operational rather than merely descriptive. They also provide concrete failure modes for evaluation: invalid resumes, stale frames, orphaned goals, silent overwrites, and unauditable handoffs.

\section{Goal Stack, Goal Tree, or Goal Graph?}

The phrase \emph{goal stack} is useful because many interruptions have call-stack semantics: the user suspends a primary goal, asks a bounded side question, and then returns to the suspended point. However, stack discipline is not universal. Table~\ref{tab:goal-structures} contrasts the main structures because real conversations often contain multiple open objectives whose relationships are not purely last-in-first-out.

\begin{table}[t]
	\centering
	\caption{Goal structures for interruptible conversational runtimes.}
	\label{tab:goal-structures}
	\small
	\begin{tabular}{p{0.22\linewidth}p{0.35\linewidth}p{0.34\linewidth}}
	\toprule
	\textbf{Structure} & \textbf{Best fit} & \textbf{Failure mode if overused} \\
	\midrule
	Goal stack & Nested interruptions with clear return points. & Forces unrelated goals into artificial push/pop order. \\
	Goal tree & Primary goals with child subgoals, explanations, policies, and clarifications. & Struggles when goals share dependencies across branches. \\
	Goal DAG & Multiple open goals with shared constraints, resources, or dependencies. & Requires stronger consistency, invalidation, and scheduling policies. \\
	\bottomrule
	\end{tabular}
\end{table}

For example, an event-registration goal may spawn venue-accessibility, catering-policy, hotel-logistics, invoice, and group-discount goals. Some are temporary side questions; others update constraints that affect the primary registration; others remain open independently. A strict stack captures the simplest case, but a tree or DAG better captures persistent subgoals, shared constraints, and non-local invalidation.

The stronger architectural claim is therefore not that all conversations should be modeled as stacks. The claim is that conversational goals should be represented explicitly, and that the runtime should choose an appropriate goal structure. The progression is evolutionary: a stack is sufficient while interruptions are nested and last-in-first-out; a tree becomes necessary when a primary objective decomposes into durable subgoals; a graph becomes necessary when goals share constraints, dependencies, approvals, or invalidation effects. In the rest of the paper, \emph{goal structure} refers to this general family, with goal stacks treated as the minimal implementation.

The GC-3 to GC-4 boundary is not merely a question of adding metadata to a tree. A goal tree assumes a dominant decomposition relation: each child goal belongs to one parent, and local metadata can usually be interpreted within that parent-child context. GC-4 breaks that assumption. A constraint, approval, tool result, or user revision may affect several goals across different branches, so correctness depends on non-tree edges: shared resources, dependency links, cross-goal consistency constraints, invalidation relations, and audit paths. Encoding these effects as ad hoc metadata on tree nodes would require application-specific propagation rules that are no longer local to the tree. GODR treats those propagation, consistency-checking, invalidation, and resumption rules as runtime semantics over an explicit goal graph.

\section{Proposed Direction: Goal-Oriented Dialogue Runtime}

We propose a Goal-Oriented Dialogue Runtime (GODR) as an architectural layer above agent runtimes and graph runtimes. A goal-stack dialogue manager is the simplest instance of this model, but not the only one. GODR does not replace LangGraph or other frameworks. Instead, it defines the missing runtime entities that complex conversations require.

\subsection{Architectural Layers}

A Goal-Oriented Dialogue Runtime separates three concerns that are often collapsed in modern agent examples. Figure~\ref{fig:godr-layers} shows the resulting layer boundary:

\begin{figure}[t]
\centering
\resizebox{0.80\linewidth}{!}{%
\begin{tikzpicture}[node distance=0.58cm, every node/.style={font=\sffamily}]

\node[layer_box, fill=layerGoal] (goal) {\textbf{Goal Layer}\\[-1pt]{\scriptsize lifecycle, structure, resumption, dependencies, invalidation}};
\node[layer_box, fill=layerDialogue, below=of goal] (dialogue) {\textbf{Dialogue Layer}\\[-1pt]{\scriptsize state tracking, memory, goal policy, human-in-the-loop policy, audit}};
\node[layer_box, fill=layerExec, below=of dialogue] (execution) {\textbf{Execution Layer}\\[-1pt]{\scriptsize LangGraph, ADK, CrewAI, AutoGen, Semantic Kernel, APIs, tools}};

\draw[godr_arrow] (goal) -- node[right, font=\scriptsize\bfseries] {interprets and governs} (dialogue);
\draw[godr_arrow] (dialogue) -- node[right, font=\scriptsize\bfseries] {delegates bounded work} (execution);
\draw[trace_arrow] (execution.west) to[out=180,in=180,looseness=1.2] node[left, font=\scriptsize\bfseries] {observations and tool results} (dialogue.west);
\draw[trace_arrow] (dialogue.west) to[out=180,in=180,looseness=1.2] node[left, font=\scriptsize\bfseries] {state updates} (goal.west);

\end{tikzpicture}
}
\caption{Three-layer separation for goal-oriented conversational architecture. GODR separates goal management, dialogue state and policy, and bounded execution instead of collapsing them into a single agent or graph controller.}
\label{fig:godr-layers}
\end{figure}
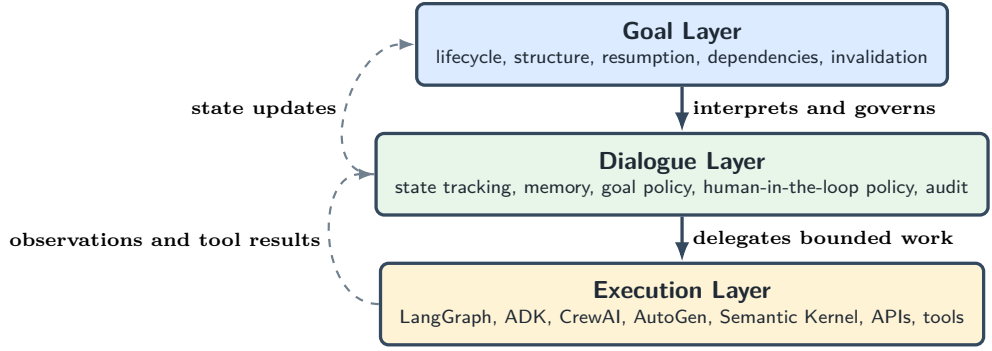

The Goal Layer owns the user's evolving objectives. The Dialogue Layer interprets turns, maintains state, applies policies, and records decisions. The Execution Layer performs bounded reasoning and tool execution. This separation allows existing orchestration frameworks to remain useful without forcing them to encode the entire conversational lifecycle.

\subsection{Goal Lifecycle}

A GODR requires an explicit lifecycle for each conversational objective. Table~\ref{tab:lifecycle} defines the states a goal can occupy, distinguishes resumable suspension from blocking or cancellation, and identifies the terminal states that close or replace an objective.

\begin{table}[t]
	\centering
	\caption{Proposed lifecycle states for interruptible conversational goals.}
	\label{tab:lifecycle}
	\small
	\begin{tabular}{p{0.22\linewidth}p{0.68\linewidth}}
	\toprule
	\textbf{State} & \textbf{Meaning} \\
	\midrule
	created & A potential goal has been detected but not yet committed. \\
	active & The goal currently controls the next conversational action. \\
	suspended & The goal is not active but has a valid resumption contract. \\
	blocked & The goal requires external input, user clarification, human review, or system recovery. \\
	completed & The goal has reached its completion criteria. \\
	abandoned & The user or policy cancelled the goal. \\
	superseded & Another goal replaced it semantically. \\
	failed & The goal cannot safely continue. \\
	\bottomrule
	\end{tabular}
\end{table}

The central distinction is between goals that can safely return to a known point and goals that require a policy decision before continuing. A suspended goal preserves a resumption contract; a blocked, failed, abandoned, or superseded goal cannot simply be resumed without additional validation.

\subsection{Core Data Structures}

The lifecycle is implemented through a small set of runtime objects. This subsection sketches the minimal data structures needed to store goal identity, task-frame state, parent-child relations, execution bindings, resumption contracts, invalidation checks, and the session-level registry that keeps these objects consistent.

\begin{lstlisting}[style=python]
class Goal:
    id: str
    type: str
    status: GoalStatus
    parent_id: str | None
    children: list[str]
    priority: int
    task_frame: dict
    local_memory: list[Event]
    active_agent: str | None
    active_graph: str | None
    resumption_contract: ResumptionContract | None
    completion_criteria: CompletionCriteria
    safety_policy: SafetyPolicy

class ResumptionContract:
    graph_id: str
    node_or_checkpoint: str
    pending_action: str
    expected_user_input: str | None
    required_context_keys: list[str]
    invalidation_conditions: list[str]

class DialogueState:
    active_goal_id: str | None
    goal_stack: list[str]
    goal_registry: dict[str, Goal]
    session_memory: dict
    global_constraints: dict
    last_user_intent: str | None
\end{lstlisting}

\subsection{Goal Operations Model and Policy}

The goal operations form a small algebra over conversational objectives. The Goal Policy is the decision layer that determines how a user utterance changes the goal structure. It is the core control point of the runtime, and should not be treated as an unconstrained LLM router. A practical policy can be implemented as a hybrid neuro-symbolic controller:

\begin{equation}
\pi(u_t,D_t)=\operatorname{argmax}_{o \in \mathcal{O}_{valid}} score(o,u_t,D_t),
\end{equation}

where $\mathcal{O}_{valid}$ is the subset of operations allowed by lifecycle constraints, business rules, safety policies, and goal-structure invariants. The scoring function may be produced by an LLM classifier, a learned ranker, symbolic rules, or a combination of these. This formulation is compatible with learned decision models, including reinforcement-learning-based policies, where $\mathcal{O}_{valid}$ acts as a symbolic action mask that restricts candidate operations to transitions satisfying lifecycle constraints, safety guards, and conversational invariants. Crucially, symbolic constraints should filter or veto operations that would violate resumability, frame ownership, invalidation safety, or human-in-the-loop requirements.

In systems terms, the Goal Policy acts as the scheduler for conversational objectives. It determines which goal receives control on a turn, which goal is suspended, which suspended goal can be resumed, and which operation must be blocked because it would violate lifecycle invariants. This scheduling role should be separated from language understanding and from policy optimization: an LLM or learned policy may propose candidate operations, but the runtime must enforce admissibility through typed state, guards, and audit requirements.

\begin{table}[t]
	\centering
	\caption{Goal-policy operations and minimal guards.}
	\label{tab:goal-policy}
	\scriptsize
		\begin{tabularx}{\linewidth}{p{0.18\linewidth}Y Y}
	\toprule
	\textbf{Operation} & \textbf{Typical trigger} & \textbf{Required guard} \\
	\midrule
	\texttt{continue} & User answers the pending question or provides expected information. & Active goal has a valid pending action. \\
	\texttt{revise} & User changes a slot, constraint, date, destination, or preference. & Revision does not invalidate completed irreversible actions. \\
	\texttt{push} & User asks a side question or creates a child objective. & Current goal can be suspended or remain concurrently open. \\
	\texttt{switch} & User changes to a sibling or independent goal. & Target goal is active, suspended, or newly creatable. \\
	\texttt{resume} & User asks to continue a previous objective. & Resumption contract is valid under current constraints. \\
	\texttt{pop} & Goal reaches completion criteria. & Terminal state and parent/previous goal are well defined. \\
	\texttt{cancel} & User cancels one or more objectives. & Cancellation scope is unambiguous or confirmed. \\
	\texttt{escalate} & Policy, uncertainty, or risk requires human review. & Escalation target and affected goals are recorded. \\
	\bottomrule
	\end{tabularx}
\end{table}

The operations in Table~\ref{tab:goal-policy} are intentionally small: they are the primitive transformations from which higher-level conversational behavior is composed. This formulation separates interpretation from permission. An LLM may infer that the user wants to resume event registration, but the runtime must still verify that the event-registration goal has a valid resumption contract, that newer turns have not invalidated its frame, and that no human approval is pending. This prevents plausible conversational behavior from bypassing operational correctness.

\subsection{Turn-Level Algorithm}

Algorithm~\ref{alg:godr-turn} summarizes the turn-level control loop. The key point is that the goal-level transition happens before any graph runtime or tool agent is invoked.

\begin{algorithm}[t]
\caption{Goal-Oriented Turn Handling}
\label{alg:godr-turn}
\begin{algorithmic}[1]
\Require user utterance $u_t$, dialogue state $D_t$
\Ensure system response $r_t$ and updated dialogue state $D_{t+1}$
\State $I_t \gets \Call{Understand}{u_t, D_t}$
\State $o_t \gets \Call{GoalPolicy}{u_t, I_t, D_t}$
\State $D_t^{pre} \gets \Call{ApplyGoalOperation}{o_t, D_t}$
\If{$o_t = \textsc{Continue}$}
    \State $g_t \gets \Call{ActiveGoal}{D_t^{pre}}$
\ElsIf{$o_t = \textsc{Push}$}
    \State $g_{active} \gets \Call{ActiveGoal}{D_t^{pre}}$
    \State $D_t^{pre} \gets \Call{SuspendIfNeeded}{g_{active}, D_t^{pre}}$
    \State $g_t \gets \Call{CreateGoal}{o_t.goalType, D_t^{pre}}$
\ElsIf{$o_t = \textsc{Resume}$}
    \State $g_t \gets \Call{ResumeGoal}{o_t.goalId, D_t^{pre}}$
\ElsIf{$o_t = \textsc{Pop}$}
    \State $D_t^{pre} \gets \Call{CompleteActiveGoal}{D_t^{pre}}$
    \State $g_t \gets \Call{SelectReturnGoal}{D_t^{pre}}$
\ElsIf{$o_t = \textsc{Cancel}$}
    \State $D_t^{pre} \gets \Call{CancelGoals}{o_t.scope, D_t^{pre}}$
    \State $g_t \gets \Call{SelectRecoveryGoal}{D_t^{pre}}$
\Else
    \State $g_t \gets \Call{EscalateOrClarify}{o_t, D_t^{pre}}$
\EndIf
\State $y_t \gets \Call{RunExecutionRuntime}{g_t, u_t, D_t^{pre}}$
\State $D_{t+1} \gets \Call{UpdateGoalAndDialogueState}{y_t, D_t^{pre}}$
\State $r_t \gets \Call{GenerateResponse}{y_t, D_{t+1}}$
\State \Return $(r_t, D_{t+1})$
\end{algorithmic}
\end{algorithm}

\section{Design Principles}

A Goal-Oriented Dialogue Runtime should follow seven design principles:

\begin{enumerate}[leftmargin=*]
\item Separate agent identity from goal identity. The active agent may change often; the active goal must remain stable unless explicitly revised.
\item Make interruption explicit. A side question should not silently overwrite the main goal state.
\item Persist resumption contracts, not only chat history. A resumable system needs a compact representation of where and how to continue.
\item Use graphs for bounded execution, not as the only representation of the whole conversation.
\item Treat state as a schema, not a dictionary. Keys need ownership, lifecycle, validation, and migration rules.
\item Keep human-in-the-loop decisions at the goal-policy level when they affect business decisions, and inside subgraphs when they affect local tool execution.
\item Prefer semantic transitions over complete edge enumeration. The policy should classify goal operations, not maintain $N^2$ edges.
\end{enumerate}

GODR should also have a clear non-use criterion. It should not be introduced when a single workflow graph already owns the user journey, interruptions are shallow, and resumption does not require goal-level auditability. In those cases, conventional workflow orchestration, root graphs, or simple dialogue stacks are usually easier to implement and maintain.

\section{Evaluation Protocol}

The central empirical hypothesis is that explicit goal-oriented runtime management improves robustness and maintainability in multi-domain conversations with interruptions, without replacing graph runtimes for bounded task execution. A strong evaluation should compare four systems under the same domains, tools, and language model: (i) a flat finite-state or router baseline, (ii) a root graph with subgraphs, (iii) a supervisor-agent architecture with shared state, and (iv) the proposed Goal-Oriented Dialogue Runtime layered above the same graph or agent runtime.

The benchmark should contain scripted and adversarial multi-objective dialogues across at least three domains. Each dialogue should include a primary goal, one or more side goals, interruptions, corrections, cancellations, resumptions, and invalidating events. For example, an event-registration dialogue may suspend registration for venue accessibility, resume the registration, revise the session date, ask about dietary policy, invalidate an earlier seat hold, and then request an invoice. Gold annotations should include active goal, suspended goals, task-frame values, expected resumption point, terminal goal statuses, and human-in-the-loop decisions when present.

The evaluation section should therefore be read as an experimental protocol for future implementations rather than as empirical evidence for performance claims. A minimal protocol would fix the language model, tools, domain APIs, and user scripts across all baselines, then vary only the dialogue-control architecture. Each run should emit a machine-readable trace of goal operations, task-frame updates, tool calls, and final responses so that conversational continuity can be scored independently of fluency.

A future benchmark, which we call \emph{GoalBench}, should isolate goal-continuity failures rather than general language quality. It can be organized around the five scenario families listed in Table~\ref{tab:goalbench-scenarios}:

\begin{enumerate}[leftmargin=*]
\item \textbf{Single interruption.} A primary goal is suspended by one side question and must resume at the exact pending action.
\item \textbf{Nested interruption.} A side goal is itself interrupted before the original goal resumes.
\item \textbf{Goal revision.} The user changes a constraint, slot, or preference while preserving the broader objective.
\item \textbf{Goal invalidation.} A later utterance or tool result invalidates a suspended goal's resumption contract.
\item \textbf{Concurrent goals.} Multiple goals remain open and must be scheduled, updated, or closed without silent overwrites.
\end{enumerate}

GoalBench would allow systems to be compared by goal-state accuracy, resumption correctness, invalid resume rate, and audit completeness, independently of surface response fluency. Table~\ref{tab:evaluation} maps these continuity failures to measurable evaluation dimensions.

\begin{table}[t]
\centering
\caption{Minimal reproducible scenarios for evaluating goal continuity.}
\label{tab:goalbench-scenarios}
\scriptsize
\begin{tabularx}{\linewidth}{p{0.18\linewidth}Y Y p{0.15\linewidth}}
\toprule
\textbf{Scenario} & \textbf{Interruption pattern} & \textbf{Expected invariant} & \textbf{Failure} \\
\midrule
Event registration & Venue, dietary, and invoice side goals interrupt session selection. & Registration resumes at the missing session slot with prior constraints preserved. & Lost goal or stale frame. \\
Procurement approval & Budget, vendor, and policy goals remain open with dependencies. & Purchase request cannot resume after a blocking approval is invalidated. & Invalid resume. \\
Travel booking & Hotel and flight goals share dates and destination constraints. & Revising dates updates dependent frames or marks them stale. & Silent overwrite. \\
Customer support & Troubleshooting is interrupted by warranty, account, and escalation checks. & Escalation preserves the failed diagnostic step and audit trail. & Missing handoff log. \\
Personal assistant & Calendar, reminder, and email goals are interleaved. & Completed and suspended goals remain distinguishable after topic switches. & Status confusion. \\
\bottomrule
\end{tabularx}
\end{table}

\begin{table}[t]
	\centering
	\caption{Evaluation dimensions and directional targets for Goal-Oriented Dialogue Runtimes.}
	\label{tab:evaluation}
	\scriptsize
	\begin{tabularx}{\linewidth}{p{0.25\linewidth}Y p{0.09\linewidth}}
	\toprule
	\textbf{Dimension} & \textbf{Metric or evidence} & \textbf{Target} \\
	\midrule
	Resumption robustness & Successful return to the correct pending action after side goals and interruptions. & Higher \\
	Goal-state accuracy & Agreement with gold active, suspended, completed, abandoned, and superseded labels. & Higher \\
	Frame consistency & Slot and constraint correctness after interruptions, revisions, and invalidations. & Higher \\
	Invalid resume rate & Attempts to resume stale, contradicted, or policy-invalid goals. & Lower \\
	Transition complexity & Number of explicit graph edges or routing rules needed to add a new domain. & Lower \\
	Audit completeness & Fraction of goal status changes and resumption-contract updates reconstructable from logs. & Higher \\
	Task success & Completion rate for primary and secondary goals under identical tool availability. & Higher \\
	\bottomrule
	\end{tabularx}
\end{table}

This protocol deliberately separates conversational correctness from model fluency. The key question is not whether an LLM can produce a plausible next answer, but whether the runtime preserves the correct goal structure over long, interruptible interactions.

\section{Research Agenda}

The proposed architecture leads to a focused research agenda:

\begin{itemize}[leftmargin=*]
\item Can goal-oriented runtime management reduce transition explosion compared with graph-only orchestration in multi-domain conversations?
\item What is the minimal set of lifecycle states needed to represent interruptible conversational goals?
\item When should an utterance create a child goal, switch to a sibling goal, revise the current goal, or resume a previous one?
\item How can resumption contracts be made framework-neutral across LangGraph, ADK, Semantic Kernel, CrewAI, and other runtimes?
\item Can LLMs reliably classify goal operations, or is a hybrid symbolic/neural policy required?
\item How should human-in-the-loop decisions be represented when they affect suspended goals?
\end{itemize}

\section{Reference Implementation Architecture}

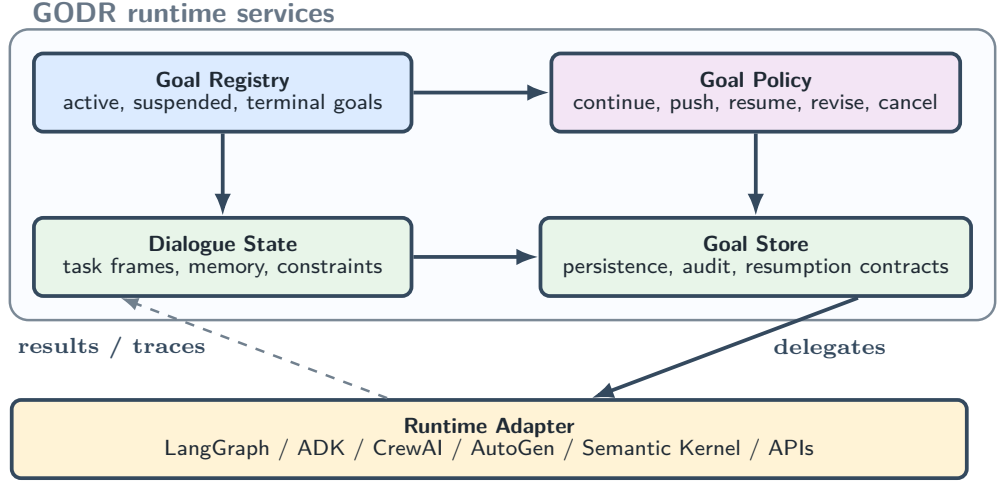
\begin{figure}[t]
\centering
\resizebox{0.80\linewidth}{!}{%
\begin{tikzpicture}[
	every node/.style={font=\sffamily},
	arch_box/.style={layer_box, minimum width=4.25cm, minimum height=0.88cm, font=\sffamily\scriptsize},
	adapter_box/.style={layer_box, fill=layerExec, minimum width=10.75cm, minimum height=0.88cm, font=\sffamily\scriptsize},
	runtime_shell/.style={draw=lineDark!65, rounded corners=6pt, thick, fill=layerGoal!13},
	arch_label/.style={fill=white, inner xsep=3pt, inner ysep=1pt, font=\scriptsize\bfseries, text=lineDark}
]

\node[arch_box, fill=layerGoal] (registry) at (-3.0,1.75) {\textbf{Goal Registry}\\[-1pt]{\scriptsize active, suspended, terminal goals}};
\node[arch_box, fill=layerPolicy] (policy) at (3.0,1.75) {\textbf{Goal Policy}\\[-1pt]{\scriptsize continue, push, resume, revise, cancel}};
\node[arch_box, fill=layerDialogue] (state) at (-3.0,-0.10) {\textbf{Dialogue State}\\[-1pt]{\scriptsize task frames, memory, constraints}};
\node[arch_box, fill=layerDialogue] (store) at (3.0,-0.10) {\textbf{Goal Store}\\[-1pt]{\scriptsize persistence, audit, resumption contracts}};
\begin{scope}[on background layer]
\node[runtime_shell, fit=(registry)(policy)(state)(store), inner sep=0.25cm] (godrshell) {};
\end{scope}
\node[anchor=west, font=\sffamily\small\bfseries, text=lineDark!75] at ([xshift=0.14cm,yshift=0.18cm]godrshell.north west) {\textbf{GODR runtime services}};

\node[adapter_box] (adapter) at (0,-2.15) {\textbf{Runtime Adapter}\\[-1pt]{\scriptsize LangGraph / ADK / CrewAI / AutoGen / Semantic Kernel / APIs}};

\draw[godr_arrow] (registry.east) -- (policy.west);
\draw[godr_arrow] (registry.south) -- (state.north);
\draw[godr_arrow] (policy.south) -- (store.north);
\draw[godr_arrow] (state.east) -- (store.west);
\draw[godr_arrow] ([xshift=1.15cm]store.south) -- node[arch_label, right=12pt] {delegates} ([xshift=1.15cm]adapter.north);
\draw[trace_arrow] ([xshift=-1.15cm]adapter.north) -- node[arch_label, left=12pt] {results / traces} ([xshift=-1.15cm]state.south);

\end{tikzpicture}
}
\caption{Reference implementation architecture for GODR. The runtime owns goal registry, policy, state, persistence, audit, and resumption contracts while delegating bounded execution through adapters.}
\label{fig:reference-architecture}
\end{figure}

A reference implementation does not require replacing the execution framework. It can be built as a thin runtime layer with explicit ownership of goal state and adapter-based delegation to existing graph or agent systems. Under this interpretation, the execution runtime performs bounded computation, while GODR provides operating-system-like services for conversational objectives~\cite{garraldabarrio2026governedevolutionagentruntimes}: registry, scheduling, persistence, context switching, and audit. Figure~\ref{fig:reference-architecture} maps these services to implementation components.

The main implementation boundary is ownership: the Goal Store owns durable goal state and resumption contracts; the execution runtime owns local graph checkpoints, tool calls, and domain workflows. This prevents duplicated state while allowing existing frameworks to remain the execution substrate.

\section{Minimal Viable Architecture}

Before building a complex framework, a pragmatic prototype can be implemented with five components:

\begin{enumerate}[leftmargin=*]
\item A typed \texttt{DialogueState} object stored per session or thread.
\item A goal registry with active, suspended, completed, and abandoned goals.
\item A small \texttt{GoalPolicy} classifier that emits operations: \texttt{continue}, \texttt{push}, \texttt{pop}, \texttt{resume}, \texttt{revise}, \texttt{cancel}, and \texttt{escalate}.
\item A \texttt{GraphRuntime} adapter that invokes LangGraph subgraphs or other agents using the selected goal context.
\item An audit log that records every goal operation and resumption contract change.
\end{enumerate}

The minimal prototype can be specified as a thin middleware loop rather than a new agent framework. It intercepts each user turn, updates the goal registry, selects a goal operation, delegates bounded execution to the selected runtime adapter, and persists both the resulting task-frame updates and the goal-level transition. Table~\ref{tab:minimal-contract} states the corresponding implementation contract.

\begin{table}[t]
\centering
\caption{Minimal implementation contract for a Goal-Oriented Dialogue Runtime.}
\label{tab:minimal-contract}
\scriptsize
\begin{tabularx}{\linewidth}{p{0.16\linewidth}Y Y Y}
\toprule
\textbf{Component} & \textbf{Responsibility} & \textbf{Input / output} & \textbf{Owned state} \\
\midrule
\texttt{DialogueState} & Carries session context, global constraints, and visible history. & User turn and runtime observations / updated session state. & History, constraints, shared context. \\
\texttt{GoalRegistry} & Maintains active, suspended, terminal, and blocked goals. & Goal operations / updated goal structure. & Goal identifiers, statuses, parent and dependency links. \\
\texttt{GoalPolicy} & Selects lifecycle operations under symbolic guards. & User turn plus current state / operation such as \texttt{push} or \texttt{resume}. & Policy rules, classifiers, validation constraints. \\
\texttt{RuntimeAdapter} & Delegates bounded work to graphs, tools, or agents. & Selected goal context / tool results and local execution traces. & Adapter configuration, not durable goal state. \\
\texttt{AuditLog} & Records goal-level decisions and resumption changes. & Lifecycle transition events / reconstructable trace. & Operation and resumption history. \\
\bottomrule
\end{tabularx}
\end{table}

The corresponding implementation sketch is deliberately small. A prototype can treat GODR as middleware around an existing graph or agent runtime:

\begin{lstlisting}[style=python]
def handle_turn(user_turn: str, state: DialogueState) -> Response:
    intent = understand(user_turn, state.visible_history)
    operation = goal_policy.select(intent, state.goal_registry)
    state = apply_goal_operation(operation, state)

    goal = state.active_goal()
    if goal is None:
        return ask_clarification(state)

    result = runtime_adapter.invoke(
        graph_id=goal.active_graph,
        context=goal.task_frame,
        checkpoint=goal.resumption_contract.node_or_checkpoint,
    )
    state = merge_result(goal.id, result, state)
    audit_log.record(operation, goal.id, result.trace_id)
    return render_response(result, state)
\end{lstlisting}

\subsection{Example: Event Registration Assistant}

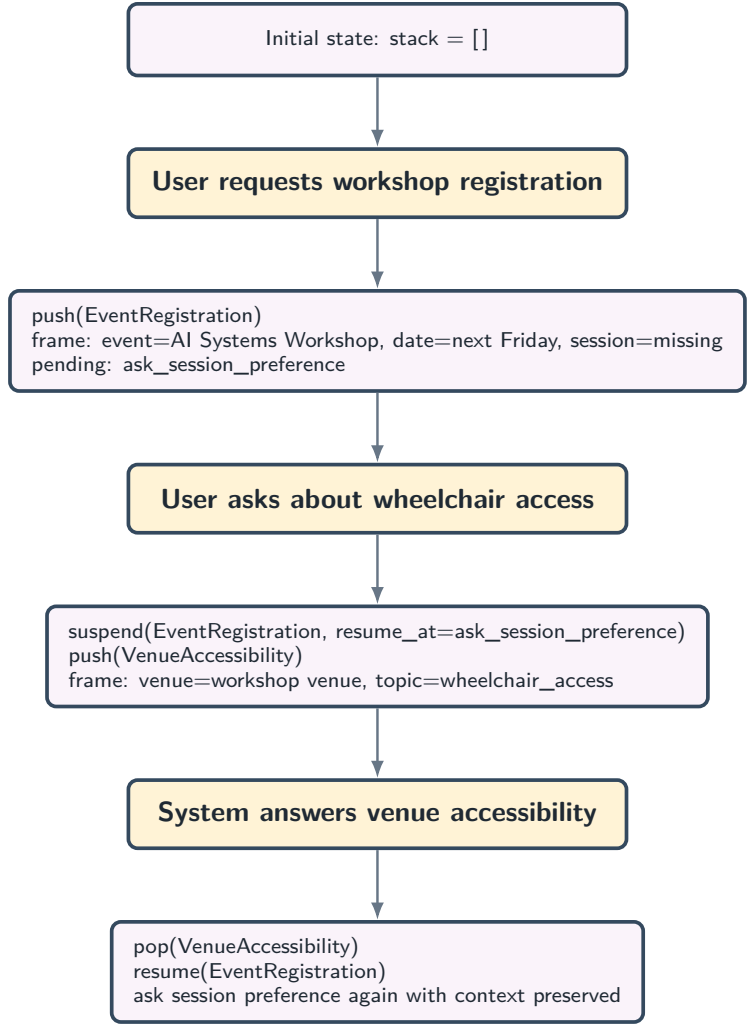
\begin{figure}[t]
\centering
\resizebox{0.60\linewidth}{!}{%
\begin{tikzpicture}[node distance=0.8cm, every node/.style={font=\sffamily}]
\node[policy_note, minimum width=5.7cm] (initial) {Initial state: stack = [\,]};
\node[workflow_root, below=of initial, minimum width=5.7cm] (register) {User requests workshop registration};
\node[policy_note, below=of register, minimum width=5.7cm] (eventframe) {push(EventRegistration)\\frame: event=AI Systems Workshop, date=next Friday, session=missing\\pending: ask\_session\_preference};
\node[workflow_root, below=of eventframe, minimum width=5.7cm] (access) {User asks about wheelchair access};
\node[policy_note, below=of access, minimum width=5.7cm] (venueframe) {suspend(EventRegistration, resume\_at=ask\_session\_preference)\\push(VenueAccessibility)\\frame: venue=workshop venue, topic=wheelchair\_access};
\node[workflow_root, below=of venueframe, minimum width=5.7cm] (answer) {System answers venue accessibility};
\node[policy_note, below=of answer, minimum width=5.7cm] (resume) {pop(VenueAccessibility)\\resume(EventRegistration)\\ask session preference again with context preserved};

\draw[workflow_arrow] (initial) -- (register);
\draw[workflow_arrow] (register) -- (eventframe);
\draw[workflow_arrow] (eventframe) -- (access);
\draw[workflow_arrow] (access) -- (venueframe);
\draw[workflow_arrow] (venueframe) -- (answer);
\draw[workflow_arrow] (answer) -- (resume);
\end{tikzpicture}
}
\caption{Trace of the event-registration example. The runtime suspends the registration goal, serves the side question, and resumes the preserved pending action.}
\label{fig:event-registration-trace}
\end{figure}

Figure~\ref{fig:event-registration-trace} revisits the event-registration example as a minimal stack trace: the runtime pushes the registration goal, suspends it for a side question, completes the side goal, and resumes the original pending action.

\subsection{Emerging GC-4 Scenarios}

The GC-4 regime is not intended to describe every chatbot or every agent workflow. It appears when a conversational system must coordinate several user objectives that remain simultaneously live and whose constraints can affect one another across workflow boundaries. A single bounded workflow is typically GC-1; a workflow with resumable side questions is often GC-2; a multi-domain assistant that tracks several independent objectives may reach GC-3; GC-4 begins when those objectives share constraints, approvals, resources, deadlines, or invalidation conditions that require runtime consistency management.

Three production-oriented scenarios illustrate the boundary. First, enterprise copilots often span procurement, approvals, vendor onboarding, budgeting, legal review, and reporting. A user may start a purchase request, ask about vendor eligibility, revise a budget constraint, and later request an invoice or approval status; these goals are not merely children of one tree, because a change in budget, vendor, or approval policy can propagate across several active objectives. Second, personal and professional digital assistants combine travel, calendar, registration, accommodation, dietary, accessibility, and expense goals. Changing a travel date may not invalidate hotel or event registration goals immediately, but it creates a cross-goal consistency obligation. Third, multi-agent business operations expose several specialized agents or workflows for customer support, finance, logistics, compliance, and human escalation. The hard problem is not only which agent acts next, but which business objective owns each partial result, which suspended objectives remain valid, and which consistency or audit checks must run before resumption.

These scenarios do not imply that all enterprise assistants require GODR. They indicate where the architectural boundary becomes visible: when correctness depends on lifecycle ownership and cross-goal consistency rather than on a richer local workflow state alone.

\subsection{Engineering Methodology}

The broader contribution of this paper is a methodology for selecting conversational architectures, not merely a proposal for one runtime. A systematic engineering process can proceed in five phases:

\begin{enumerate}[leftmargin=*]
\item \textbf{Characterize the conversation.} Identify number of objectives, domain breadth, interruption freedom, resumption requirements, user corrections, human-in-the-loop points, and dependency structure.
\item \textbf{Classify runtime complexity.} Assign the system to a low, medium, high, or very high complexity regime based on objective multiplicity, interruption depth, and dependency coupling.
\item \textbf{Select the dialogue architecture.} Choose the simplest architecture that matches the observed goal complexity: FSM or workflow graph for bounded processes, goal stack for nested resumability, agenda or goal tree for independent concurrent goals, and GODR with a goal graph for GC-4 dependency and invalidation cases.
\item \textbf{Select the execution engine.} Map bounded task execution to LangGraph, ADK, CrewAI, AutoGen, Semantic Kernel, custom workflows, or conventional service orchestration.
\item \textbf{Select the goal structure.} Use a stack for nested interruptions, a tree for decomposed objectives, or a DAG for concurrently open and interdependent goals.
\end{enumerate}

This methodology reframes the design problem. Instead of asking which agent framework should own the whole conversation, the engineer first characterizes conversational complexity, then selects the minimal runtime abstraction that preserves correctness. GODR is therefore one point in a broader design space: excessive for simple single-process workflows, optional for shallow resumability, partially useful for GC-3 agenda management, and justified when GC-4 dependencies, invalidations, and audit requirements become non-local.

\section{Contributions, Scope, and Validity}

The four contributions stated in the introduction can be grouped into two broader claims. First, the paper characterizes the design problem: multi-objective interruptible dialogue requires distinguishing active-agent selection and execution continuity from active-goal continuity, then selecting the minimal runtime abstraction for the observed goal-complexity level. This includes the taxonomic claim that stacks, trees, and goal graphs are not interchangeable implementation details; they correspond to different interruption, decomposition, dependency, and invalidation regimes. Second, the paper proposes GODR as the architecture for the GC-4 region, where goal lifecycles, goal-structure operations, task frames, interruption points, invalidation rules, resumption contracts, and audit requirements become necessary runtime concerns.

The scope is intentionally architectural and methodological. The claim is not that GODR replaces graph runtimes, agent handoffs, retrieval systems, or business workflows. The narrower claim is that when users can freely suspend, revise, supersede, and resume interdependent objectives, goal continuity should be represented explicitly rather than inferred indirectly from agent identity, chat history, memory traces, or graph position. The proposed engineering methodology follows from this boundary: characterize the conversation first, classify goal complexity second, and only then choose the dialogue architecture, execution engine, and goal structure.

Several limitations remain. Goal misclassification can corrupt intent continuity, and over-engineering is a risk for bounded workflows where root graphs plus subgraphs are sufficient. Resumption contracts require invalidation rules because a suspended goal may no longer be valid after later user actions. Framework integration must avoid duplicated ownership: GODR should own durable goal state, while graph runtimes should own local checkpoints and tool execution state.

The paper is primarily a conceptual systems paper. It does not provide a production implementation or controlled ablation study, and there is not yet a widely accepted benchmark for interruptible multi-objective conversational continuity. Existing task-oriented dialogue benchmarks emphasize slot filling, intent accuracy, or task completion more than goal suspension, resumption, cross-goal invalidation, and auditability. The proposed runtime objects and metrics should therefore be read as a design hypothesis and evaluation agenda, not as a measured performance claim. Future work should validate the separation through reference implementations, benchmark tasks, ablation studies, and longitudinal maintainability analyses.

\section{Conclusion}

This work argues for a shift in abstraction in conversational AI systems. While graph-based orchestration is highly effective for process-driven interactions, it becomes increasingly difficult to maintain conversational continuity when multiple user objectives remain active, can be suspended and resumed, share constraints, or invalidate one another.

The proposed Goal-Oriented Dialogue Runtime (GODR) reintroduces explicit goal management as a first-class runtime concern, extending principles found in mature pre-LLM dialogue systems to modern graph- and agent-based architectures. Rather than treating goals as incidental attributes of execution, GODR models them as explicit operational entities with lifecycles, ownership, interruption semantics, resumption contracts, and dependency structures.

The central architectural principle is not to replace existing execution frameworks, but to position them correctly. Graph runtimes, agents, tools, and workflow engines remain responsible for bounded task execution. GODR operates at a different level of abstraction: preserving objective continuity across interruptions, agent changes, and evolving conversational contexts.

More broadly, the paper suggests that conversational continuity should not be inferred indirectly from agents, memory traces, or execution-graph position alone. It should be represented explicitly through goal structures with lifecycle semantics. In this view, existing orchestration approaches primarily model execution continuity, whereas GODR models objective continuity.

\FloatBarrier

\section*{Acknowledgements}
The author acknowledges the Laboratorio de Innovación Aplicada (L2IA) at Minsait (Indra Group) for fostering an environment that encourages scientific exploration in AI systems, distributed runtimes, and applied agentic infrastructures.

\bibliographystyle{unsrtnat}
\bibliography{references}

\end{document}